# Interlayer magnetic interactions and ferroelectricity in $\pi/3$-twisted CrX$_2$ (X = Se, Te) bilayers

Wenqi Yang[1,2], Xinlong Yang[1,2], Menglei Li[3], Lin Hu[1,2,*] and Fawei Zheng[1,2,*]

[1] Centre for Quantum Physics, Key Laboratory of Advanced Optoelectronic Quantum Architecture and Measurement (MOE), School of Physics, Beijing Institute of Technology, Beijing 100081, China

[2] Beijing Key Lab of Nanophotonics and Ultrafine Optoelectronic Systems, School of Physics, Beijing Institute of Technology, Beijing 100081, China

[3] Department of Physics, Capital Normal University, Beijing, 100048, China

[*] Authors to whom correspondence should be addressed: hulin@bit.edu.cn and fwzheng@bit.edu.cn

**ABSTRACT**

Recently, two-dimensional (2D) bilayer magnetic systems have been widely studied. Their interlayer magnetic interactions play a vital role in the magnetic properties. In this paper, we theoretically studied the interlayer magnetic interactions, magnetic states and ferroelectricity of $\pi/3$-twisted CrX$_2$ (X = Se, Te) bilayers ($\pi/3$-CrX$_2$). Our study reveals that the lateral shift could switch the magnetic state of the $\pi/3$-CrSe$_2$ between interlayer ferromagnetic and antiferromagnetic, while just tuning the strength of the interlayer antiferromagnetic interactions in $\pi/3$-CrTe$_2$. Furthermore, the lateral shift can alter the off-plane electric polarization in both $\pi/3$-CrSe$_2$ and $\pi/3$-CrTe$_2$. These results show that stacking is an effective way to tune both the magnetic and ferroelectric properties of 1T-CrX$_2$ bilayers, making the 1T-CrX$_2$ bilayers hold promise for 2D spintronic devices.

Since graphene's successful exfoliation in 2004[1], two-dimensional (2D) materials have been extensively studied owing to their outstanding properties. Layered materials are stacked by weak van der Waals (vdW) interactions and have minor interlayer wavefunction overlaps[2], which make the systems possess interesting properties[3]. A bilayer system is composed of two monolayers, and may exhibit novel properties, such as quantum Hall effects[4], interlayer excitons[5], higher-order topological insulator[6], superconductivity[7], and abnormal conductivity[8]. The properties of a bilayer system can be effectively tuned through common methods, such as strain[3,9], magnetic field[10-12], electric field[13-15], and optics[16], and may also be tuned through the interlayer stacking[17-19], which is unique in the layered systems.

Besides the nonmagnetic bilayers, the physical properties of magnetic bilayers can also be tuned by the interlayer stacking. There are many works focusing on the stacking tuning method in bilayer magnetic systems. For example, the interlayer magnetic order in $CrI_3$ depends on the stacking configuration. The correlation between stacking structures and magnetism in bilayer $CrI_3$ is calculated, and concluded that the stacking-dependent magnetism mediated by the I octahedron exists in a wide range of 2D honeycomb magnets. These predicted results are also confirmed by linear magnetoelectric effect measurements[17]. Meanwhile, there is a study which investigates the variation of interlayer magnetic coupling in π/3-twisted bilayer $CrI_3$ with different stacking structures. It is indicated that the interlayer antiferromagnetic interaction of $\overline{A}$A-stacking configuration is very sensitive to external pressure[21]. This suggests the potential application of π/3-$CrI_3$ as a pressure sensor, and an ordered distribution of skyrmions was found in this system[22].

Here, we focused on the bilayer $CrX_2$ (X = Se, Te). The ultra-thin $CrSe_2$ layers were successfully fabricated in experiment recently. They are strong ferromagnetic and have excellent stability[22,23]. Magnetic transport studies conducted with standard Hall rod devices have revealed that $CrSe_2$ exhibits a distinct anomalous Hall effect with a Curie temperature of approximately 110 K and no obvious change after being exposed to air for several months[22]. Additionally, several approaches, such as the mechanical exfoliation method[24-26], chemical vapor deposition[27], and molecular beam epitaxy[28-30], have been employed to synthesize few-layer $CrTe_2$. $CrTe_2$ monolayer[28,31] has strong magnetic anisotropy and high Curie temperature ($T_c$ = 300 K). It has attracted wide attention in magnetic sensors and storage devices. These findings indicate that both $CrSe_2$ and $CrTe_2$ ultrathin films hold great potential for spintronics applications. However, the investigation of interlayer interactions in a π/3-$CrX_2$ remains lacking.

In this work, we systematically studied the interlayer magnetic interactions of the π/3-twisted bilayer $CrX_2$ (π/3-$CrX_2$) by using first-principles calculations. Our results reveal that the interlayer magnetic interactions of the π/3-$CrSe_2$ can be effectively tuned to ferromagnetic by lateral shift. However, the lateral shift does not affect the interlayer antiferromagnetism of the π/3-$CrTe_2$. We also found the ferroelectricity in the out-of-plane can be realized by lateral shifting in both π/3-$CrSe_2$ and π/3-$CrTe_2$.

These results show that π/3-CrX$_2$ bilayers hold promise for 2D spintronic devices.

The first-principles calculations in this work were performed by using the Vienna *ab initio* simulation package (VASP)[32]. The generalized gradient approximation (GGA) implemented as the Perdew-Burke-Ernzerhof (PBE) form was used for the exchange-correlation functional[33]. The cutoff energy was set to be 520 eV. A thickness over 15 Å was set for the vacuum space to exclude interlayer interactions, and the DFT-D3 method[34] was considered for the vdW corrections. The DFT + U method was used to account for the Coulomb correlation, and the U values were set to be 2 and 4 eV for the Cr atoms following Li's and Wu's works[35,23]. The structural relaxation was performed to ensure that the residual force on each atom and convergence criteria for the energy of the self-consistent field were less than 0.2 meV/Å and $1.0 \times 10^{-7}$ eV, respectively. A 15 × 15 × 1 *k*-mesh with the Monkhorst-Pack method was employed for Brillouin zones integration[36]. The optimal lattice parameters are a = 3.63 Å in bilayer CrSe$_2$ and a = 3.81 Å in bilayer CrTe$_2$, which are in good agreement with experimental results[22,28]. The spin exchange parameters were obtained by using the OpenMX package[37]. We also calculate the interlayer hopping parameters using the Wannier90 code[38]. The OpenMX code utilizes norm-conserving pseudopotentials[39] and a localized basis set[40]. It has been successfully employed in the study of magnetic materials[41,42]. The code is capable of producing spin exchange parameters that are qualitatively consistent with the magnetic properties obtained using the plane wave code[43].

As shown in Fig. 1, each Cr atom in the CrX$_2$ monolayer is at the center of the octahedron formed by the nearest Se/Te atoms. The ground states of monolayer CrSe$_2$ and CrTe$_2$ are ferromagnetic (FM) and can be tuned by strain, which is the result of the competition between the antiferromagnetic (AFM) direct exchange interaction between Cr-Cr and the FM superexchange interaction along Cr-Se/Te-Cr[23]. The π/3-CrX$_2$ is composed of a normal CrX$_2$ bottom layer and a π/3-rotated CrX$_2$ top layer. Three typical stacking structures, $\overline{A}A$, $\overline{A}B$ and $\overline{A}C$, are shown in Fig. 1. The $\overline{A}A$ stacking configuration has no lateral shift between the two layers, and the $\overline{A}B$ and $\overline{A}C$ stacking configurations correspond to the 2/3 and 1/3 fractional lateral shift in the $[1\overline{1}0]$ direction, respectively.

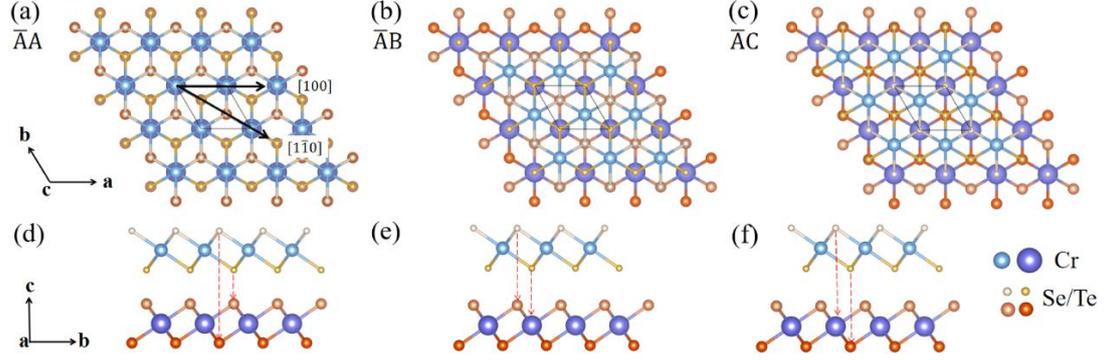

**FIG. 1.** The atomic structures for different stacking configurations of $\pi/3$-CrX$_2$. (a-c) Top view and (d-f) side view of $\pi/3$-CrX$_2$ in $\bar{A}A$- (a and d), $\bar{A}B$- (b and e) and $\bar{A}C$-stacking (c and f). The solid black lines indicate the unit cells, and the high-symmetry [100]- and [1$\bar{1}$0]-lateral shift directions are labeled in black arrows. The Cr and Se/Te atoms are shown in different colors and sizes for different layers.

More stacking structures can be constructed by moving the top layer along [100] and [1$\bar{1}$0] directions. Let's first set U = 4 eV in the DFT calculations for $\pi/3$-CrSe$_2$, according to Wu's work[23]. As shown in Fig. 2(a), the energy of the $\bar{A}A$-stacking was set to zero as a reference. In the [1$\bar{1}$0] direction, the stacking energy gets maximum at $\bar{A}B$ and $\bar{A}C$ stacking configurations. Meanwhile, in the [100] direction, there is only one stacking energy maximum, which is located at the midpoint. To see the landscape of stacking energy for full 2D lateral shifts, a 10 × 10 grid of full 2D shift was performed, as shown in Fig. 2(c). These results verify that the configurations with the stacking energy maximum in the whole 2D shift are $\bar{A}B$ and $\bar{A}C$ stacking configurations.

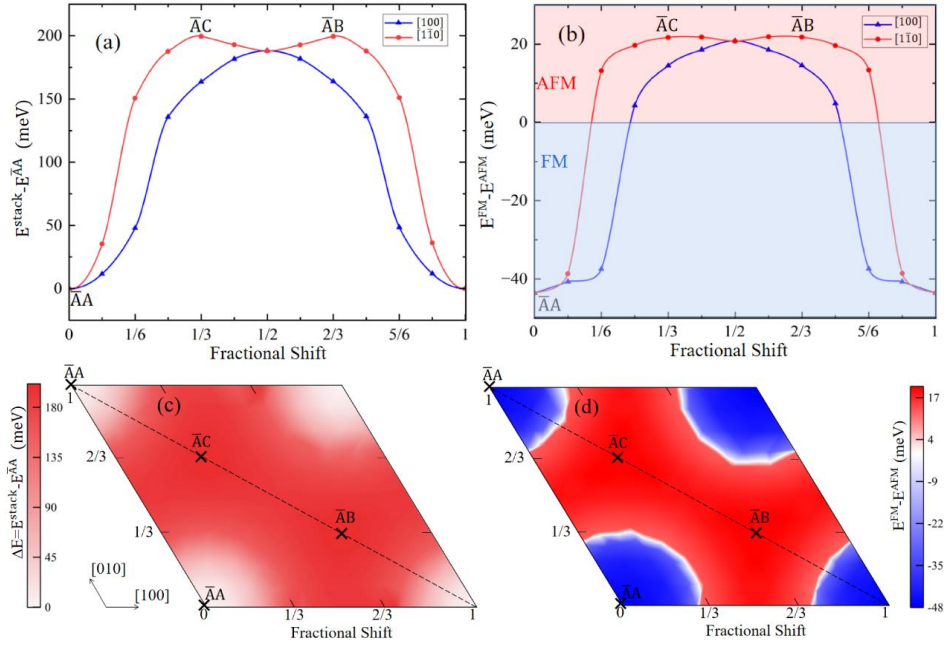

**FIG. 2.** The stacking energy and the interlayer exchange energy as a function of lateral shift in π/3-CrSe$_2$. (a-b) The stacking energy and interlayer exchange energy for lateral shifts along [100] and [1$\bar{1}$0] high symmetry lines, respectively. (c-d) The stacking energy and interlayer exchange energy for full 2D lateral shifts. The Hubbard-U value was set to 4 eV.

The π/3-CrSe$_2$ exhibits either FM or AFM interlayer interactions depending on lateral shift. The interlayer exchange energy is defined as the energy difference between FM and AFM interlayer configurations. The significant correlation between the stacking order and the magnetism is shown in Fig. 2(b). We found that $\bar{A}A$-stacking strongly prefers FM, the corresponding interlayer exchange energy is about -43.6 meV. Meanwhile, $\bar{A}B$ and $\bar{A}C$ stacking configurations prefer AFM in the [1$\bar{1}$0] direction and the associated value is as high as 21.8 meV, indicating a strong AFM interlayer interaction. Fig. 2(d) depicts the landscape of interlayer exchange energy for full 2D lateral shifts. These results indicate that the interlayer exchange energy of $\bar{A}B$ and $\bar{A}C$ stacking configurations remains the global maximum. Additionally, the case when the Hubbard-U value equals 2 eV is also considered. It can be observed that the stacking energy and interlayer exchange energy exhibit the same trend, while the values of stacking energy and exchange energy are significantly reduced. These results indicate that the Hubbard-U value does not qualitatively alter the properties of the π/3-CrSe$_2$ system during lateral shift. More detailed results can be found in the Supplementary Material (SM).

Here, the Hamiltonian is constructed to describe the magnetic interactions of the

π/3-CrSe$_2$. The Heisenberg Hamiltonian for the π/3-CrSe$_2$ can be written as:

$$H = H_i + H_o,$$

$$H_i = -\sum_{i,j} J_{i,j} S_i^t \cdot S_j^t - \sum_{i,j} J_{i,j} S_i^b \cdot S_j^b,$$

$$H_o = -\sum_{i,j} J_{i,j} S_i^t \cdot S_j^b = -J_{z1} \sum_{\langle ij \rangle} S_i^t \cdot S_j^b - J_{z2} \sum_{\langle\langle ij \rangle\rangle} S_i^t \cdot S_j^b + \cdots \quad (1)$$

Where $H_i$ and $H_o$ represent the intralayer interaction and the coupling between the two layers, respectively. The $S_i^t$ and $S_j^b$ are the magnetic moments of the $i$-th and $j$-th Cr atoms for the top and bottom CrSe$_2$ layers, respectively. The magnetic moments on Cr atoms are all normalized to 1. The $\langle ij \rangle$ and $\langle\langle ij \rangle\rangle$ are the nearest and the next-nearest interlayer neighbors of Cr atoms, and the parameters are $J_{z1}$ and $J_{z2}$, as shown in Fig. 3(a). The values of nearest- (1nn), next-nearest- (2nn) and third-nearest-neighbor (3nn) interlayer magnetic interactions in the $\overline{A}A$-stacking bilayer CrSe$_2$ can be found in Table I. As shown in Fig. 3(b), we can see that the magnetic interactions decrease quickly with increasing the Cr-Cr distances. The interaction of 3nn is already very weak. We find that the 1nn exchange parameter of $\overline{A}A$-stacking is very large, up to 18.05 meV, which indicates a strong FM interaction. To find the origin of this strong FM interlayer interaction of the $\overline{A}A$-stacking bilayer CrSe$_2$, we further listed the interlayer distances between Se atoms (Table I). The distance of the $\overline{A}A$-stacking is the shortest, indicating a very strong interlayer interaction.

TABLE I. The exchange coupling parameters of Cr atoms (meV) and the distance of Se-Se (Å) between the top and bottom CrSe$_2$ layers in $\overline{A}A$, $\overline{A}B$ and $\overline{A}C$ stacking configurations.

| neighbor | $\overline{A}A$ | $\overline{A}B$ | $\overline{A}C$ |
|---|---|---|---|
| 1nn | 18.05 | -0.71 | -0.71 |
| 2nn | -1.36 | 0.03 | 0.03 |
| 3nn | 0.60 | 0.01 | 0.01 |
| distance | $\overline{A}A$ | $\overline{A}B$ | $\overline{A}C$ |
| Se-Se | 2.46 | 3.37 | 3.36 |

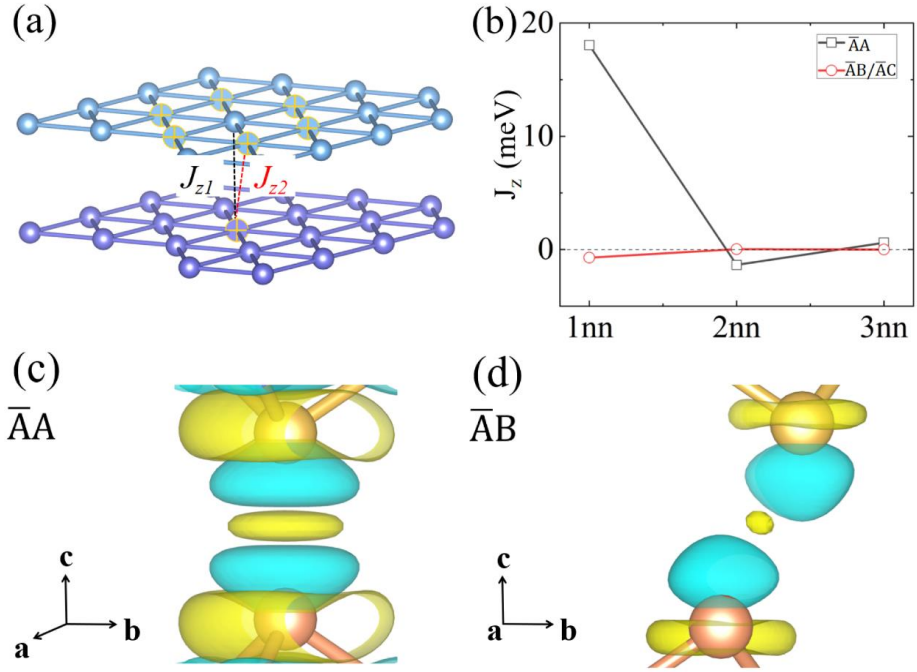

**FIG. 3.** (a) Exchange parameters are marked with dotted lines connected between Cr atoms in the $\bar{A}A$-stacking bilayer CrSe$_2$. The first ($J_{z1}$) and the second neighboring interlayer couplings ($J_{z2}$) are represented by black and red colors, respectively. The blue and purple atoms represent Cr atoms in the top and bottom CrSe$_2$ layers, respectively. (b) The exchange parameters of $\bar{A}A$ and $\bar{A}B/\bar{A}C$ stacking configurations as a function of the neighboring distance. The positive and negative values correspond to FM and AFM interactions, respectively. The charge density difference diagram of the $\bar{A}A$-stacking (c) and $\bar{A}B$-stacking (d). For clarity, this only shows the charge transfer near the Se-Se bond. Yellow represents a decrease in charge density and cyan represents an increase in charge density.

Notably, the $\bar{A}A$-stacking has the largest hopping parameter 2.51 eV between the interlayer Se atoms. However, these values in both $\bar{A}B$- and $\bar{A}C$-stacking configurations are 0.5 eV and 0.5 eV, respectively. The differential charge between the interlayer Se atoms is shown in Fig. 3(c). It can be found that there is electron accumulation between Se atoms in both $\bar{A}A$ and $\bar{A}B$ stackings, indicating the forming of a chemical bond. The electron accumulation in $\bar{A}A$ stacking is much more than that in $\bar{A}B$ stacking, which explains the much stronger interlayer hopping parameters. These results show that the strong interlayer FM of $\bar{A}A$-stacking is related to the strong interactions between Se atoms across the layer.

Besides the π/3-CrSe$_2$, the energy changing of the lateral shift in the π/3-CrTe$_2$ was also considered. Firstly, the Hubbard U value was set to 2 eV according to Li's work[35].

As shown in Fig. 4(a), there are two minimums of the stacking energy in the $[1\bar{1}0]$ direction, which are $\overline{A}B$ and $\overline{A}C$ stacking configurations. The largest stacking energy appears at the $\overline{A}A$-stacking. The minimum of stacking energy in the [100] direction is located at 1/2 lateral shift. The stacking energy of a 10 × 10 grid for the shift was also calculated. From the results, we can see that the stacking energy of $\overline{A}B$ and $\overline{A}C$ stacking configurations in the whole 2D shift are still the lowest. These results show that the stacking energy changing of the lateral shift in the π/3-CrTe$_2$ is quite different to the that of the π/3-CrSe$_2$. The $\overline{A}A$-stacking is the most stable configuration of the π/3-CrSe$_2$ during the lateral shift in both $[1\bar{1}0]$ and [100] directions. Conversely, the $\overline{A}A$ stacking configuration of the π/3-CrTe$_2$ is the most energetically unfavorable, due to the lack of strong chemical bond between interlayer Te atoms.

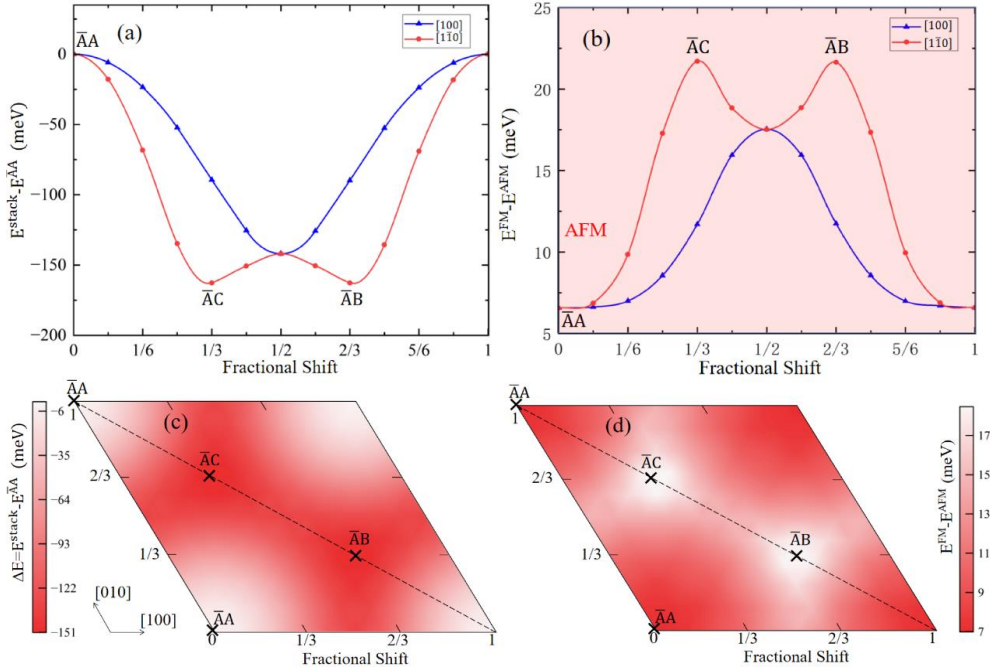

**FIG. 4.** The stacking energy and the interlayer exchange energy as a function of lateral shift in π/3-CrTe$_2$. (a-b) The stacking energy and interlayer exchange energy for lateral shifts along [100] and $[1\bar{1}0]$ high symmetry lines, respectively. (c-d) The stacking energy and interlayer exchange energy for full 2D lateral shifts. The Hubbard-U value was set to 2 eV.

The interlayer exchange energy of the **π/3**-CrTe$_2$ is also investigated. Notably, as shown in Fig. 4(b), all the interlayer exchange energy in the [100] and $[1\bar{1}0]$ directions are above zero, indicating that all configurations have AFM order. Along the [100] direction, the maximum of the interlayer exchange energy (~17.5 meV) is located at 1/2 lateral shift. While in the $[1\bar{1}0]$ direction, the maximum of the interlayer exchange energy (~21.7 meV) is located at $\overline{A}B$ and $\overline{A}C$ stacking configurations. Fig.

4(d) depicts the landscape of interlayer exchange energy for full 2D lateral shift, and the strongest AFM interlayer interactions still appear in $\overline{AB}$ and $\overline{AC}$ stacking configurations. These results reveal that the lateral shift cannot alter the magnetic ground state of the $\pi/3$-CrTe$_2$. In the $\overline{AA}$ stacking of $\pi/3$-CrTe$_2$, the interlayer distance is 3.26 Å, which is larger than that of $\pi/3$-CrSe$_2$. However, under a much smaller interlayer distance of 2.46 Å, the interlayer exchange energy will become to -89.94 meV, indicating a strong FM interaction. This numerical evidence demonstrates the interlayer distance play a critical role in interlayer exchange interactions. We also calculated the interlayer interactions with a larger Hubbard-U value (4 eV). The trend of these two kinds of energy is not changed, but the exchange strength is enhanced. The maximum of the interlayer exchange energy ~40.5 meV is at either $\overline{AB}$ or $\overline{AC}$ stacking configuration in the $[1\overline{1}0]$ direction. More details can be found in the SM.

Another interesting point we found in this $\pi/3$-CrX$_2$ system is that the ferroelectricity can be induced by lateral shifts. The monolayer CrSe$_2$ and CrTe$_2$ have inversion symmetry. According to the theory of bilayer stacking ferroelectricity[44], bilayer CrSe$_2$ and CrTe$_2$ can exhibit out-of-plane polarization by breaking the inversion symmetry. The out-of-plane polarization can be calculated by integrating the contribution from electrons and ions in the vertical plane direction, as shown in Fig. 5. In the $\pi/3$-CrSe$_2$, the $\overline{AA}$-stacking configuration exhibits no polarization due to preserved inversion symmetry. We found that the maximums of polarization were located at the $\overline{AB}$ and $\overline{AC}$ stacking configurations, with values of 4.54 × 10$^{-4}$ e/Å and -3.98 × 10$^{-4}$ e/Å, respectively. In the $\pi/3$-CrTe$_2$, the situation is almost the same as that in the $\pi/3$-CrSe$_2$, and the maximums of polarization values are 7.38 × 10$^{-4}$ e/Å and -6.94 × 10$^{-4}$ e/Å for the $\overline{AB}$ and $\overline{AC}$ stacking configurations, respectively. However, with the slip in the [100] direction, the out-of-plane polarization values are all smaller than 5.73 × 10$^{-5}$ e/Å in the $\pi/3$-CrX$_2$. As shown in Fig. 4(a) and Fig. 5(c), in the $\pi/3$-CrTe$_2$, there is no energy barrier for switching the polarization with lateral shift. However, with the slip in the $[1\overline{1}0]$ direction of the $\pi/3$-CrSe$_2$, the direction of the out-of-plane polarization is reversed by overcoming a small energy barrier of 10.32 meV/f.u. from $\overline{AC}$ to $\overline{AB}$ stacking configuration. Compared with the polarization induced in CrI$_3$[44], the value of CrX$_2$ is much smaller than that of CrI$_3$. Since CrSe$_2$ and CrTe$_2$ are metallic, the electric screen effect is significant. As a result, the polarization is smaller compared to insulating ferroelectric systems.

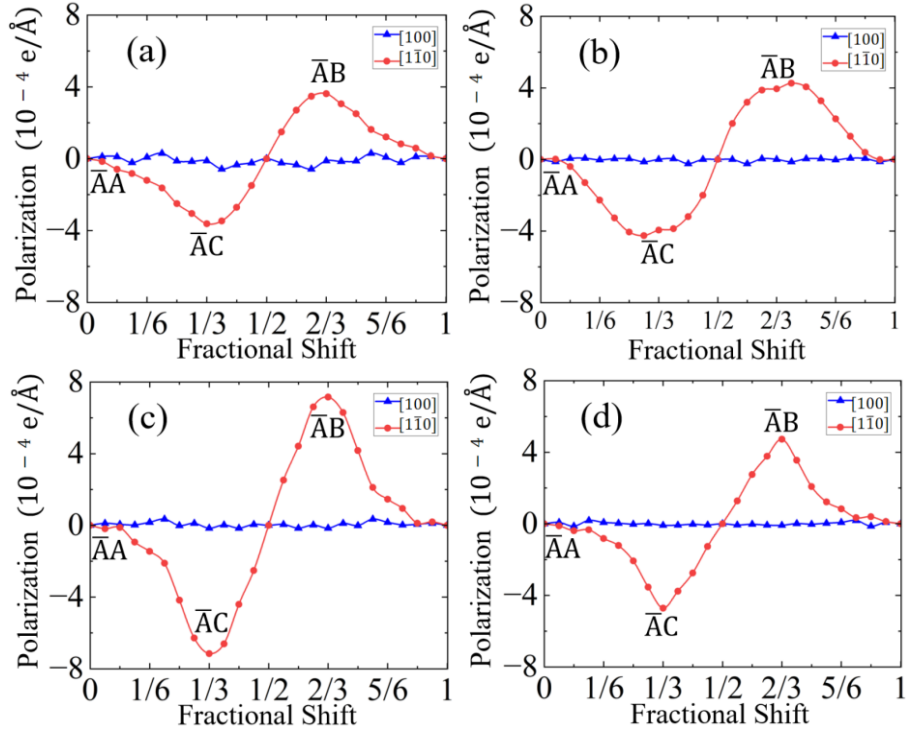

**FIG. 5.** (a-b) Dependence of polarization on shifting distance along the [100] and [1$\bar{1}$0] high symmetry lines of $\pi/3$-CrSe$_2$. (c-d) Dependence of polarization on shifting distance along the [100] and [1$\bar{1}$0] high symmetry lines of $\pi/3$-CrTe$_2$. (a) and (c) The Hubbard-U value was set to 2 eV. (b) and (d) The Hubbard-U value was set to 4 eV.

In conclusion, we have studied the interlayer interactions in lateral shifted $\pi/3$-CrX$_2$. Our study reveals that the lateral shift could effectively tune the magnetic ground state of the $\pi/3$-CrSe$_2$, while keeping the $\pi/3$-CrTe$_2$ AFM. Furthermore, the lateral shift can also alter the off-plane electric polarization in both $\pi/3$-CrSe$_2$ and $\pi/3$-CrTe$_2$. The coexistence of ferromagnetism and ferroelectricity and their tunability by lateral shift make the 1T-CrX$_2$ bilayers promising materials for 2D spintronic devices.

See the supplementary material for further details about the interlayer interactions in $\pi/3$-CrSe$_2$ for U = 2 eV and $\pi/3$-CrTe$_2$ for U = 4 eV.

This work was financially supported by National Natural Science Foundation of China (Grants Nos. 12022415, 11974056, 12274024). L.H. also thanks the support of National Key Research and Development Program of China (Grant No. 2022YFA1405600) and the Beijing Institute of Technology Research Fund Program for Young Scholars. We also acknowledge the computing resources of HPC clusters at BIT and the Shenzhen Cloud Computing Center.

# AUTHOR DECLARATIONS

**Conflict of Interest**

We declare no conflict of interest.

# DATA AVAILABILITY

The data that support the findings of this study are available within the article and its supplementary material.